\PassOptionsToPackage{table,xcdraw}{xcolor}

\documentclass[10pt, conference, compsocconf]{IEEEtran}

\usepackage{listings}
\usepackage{multirow}
\usepackage{xspace}
\usepackage{booktabs} 
\usepackage{setspace}
\usepackage{todonotes}
\usepackage{enumitem}
\usepackage{graphicx}
\usepackage{url}
\usepackage{amsfonts}
\usepackage{amssymb}
\usepackage[bookmarks=false]{hyperref}

\newcommand{\CBR}{\textit{CBR}\xspace}

\newcommand{\FSA}{\textit{FSA}\xspace}

\newcommand{\Controlled}{\textit{Controlled Burst Recording}\xspace}

\newenvironment{mod}{\color{black}}{\color{black}}

\newcommand{\CODE}[5]{
\lstinputlisting[
     caption={#2},%
     label={#1},%
     basicstyle=\scriptsize\ttfamily,%
     showstringspaces=false,%
     frame={tb},%
     lineskip=-0.4pt,%
     extendedchars=true,%
     numbers=left,%
     numbersep=4pt, 
     stepnumber=1,%
     keepspaces=true,
     firstnumber=#4,
     numberstyle=\tiny,%
     xleftmargin=\parindent,%
     breaklines=true,
     language=#5,%
     tabsize=2,
     captionpos=b,
    backgroundcolor=\color{white},
    morekeywords={aspect,thisJoinPoint,around},
]{code/#3}
\vspace{0.3cm}

}

\newcommand{\JCODESTEP}[4]{
        \CODE{#1}{#2}{#3}{#4}{java}
}

\ifCLASSINFOpdf
\else
\fi

\IEEEoverridecommandlockouts

\begin{document}

\title{CBR: Controlled Burst Recording}

\author{\IEEEauthorblockN{Oscar Cornejo*}
\IEEEauthorblockA{SnT Centre for Security, Reliability and Trust\\
University of Luxembourg, Luxembourg\\
Email: oscar.cornejo@uni.lu
\thanks{* Part of this work was carried out while the author was affiliated with University of Milano Bicocca}}
\and
\IEEEauthorblockN{Daniela Briola, Daniela Micucci and Leonardo Mariani}
\IEEEauthorblockA{Department of Informatics, Systems and Communication\\
University of Milano Bicocca, Milan 20126, Italy\\
Email: \{daniela.briola, daniela.micucci, leonardo.mariani\}@unimib.it}}


\maketitle

\begin{abstract}
Collecting traces from software running in the field is both useful and challenging. Traces may indeed help revealing unexpected usage scenarios, detecting and reproducing failures, and building behavioral models that reflect how the software is actually used. On the other hand, recording traces is an intrusive activity that may annoy users, negatively affecting the usability of the applications, if not properly designed.  

In this paper we address field monitoring by introducing \Controlled, a monitoring solution that can collect comprehensive runtime data without compromising the quality of the user experience. The technique encodes the knowledge extracted from the monitored application as a finite state model that both represents the sequences of operations that can be executed by the users and the corresponding internal computations that might be activated by each operation. 


\begin{mod}
  Our initial assessment with information extracted from ArgoUML shows that \Controlled can reconstruct behavioral information more effectively than competing sampling techniques, with a low impact on the system response time.  
\end{mod}

\end{abstract}


\IEEEpeerreviewmaketitle


\section{Introduction}


Field data is an essential source of information for a number of tasks, such as discovering emerging usage scenarios~\cite{Srivastava:WebUsage:SIGKDD:2000}, profiling users~\cite{Elbaum:Profiling:TSE:2005}, obtaining data about the reliability of the software~\cite{Gazzola:FieldFailures:ISSRE:2017}, mining models~\cite{ohmann2017lightweight,Mariani:Revolution:ISSRE:2012}, and validating software~\cite{Ohmann:OptimizedCoverage:ASE:2016,Pavlopoulou:ResidualCoverage:ICSE:1999}. The importance of observing the software while running in the field has been also well-recognized by industry: for instance, the video streaming company Netflix has started testing and collecting data directly from the field, using fault-injection and monitoring techniques~\cite{Basiri:Netflix:ISSRE:2016}.



Collecting data from the field can be \emph{challenging}. In particular, monitoring and data collection can easily interfere with the user activity~\cite{Orso:DeployedSoftware:FoSER:2010}. 
While some solutions, such as simple crash reporting features, require collecting relatively few data for a short amount of time (e.g., a snapshot of the system at the time of the crash~\cite{Eclipse:website:2018,Windows:website:2019,Delgado:TaxonomyFaultMonitoring:TSE:2004}), many interesting and sophisticated approaches require monitoring applications more extensively. 
For example, many approaches collect \emph{sequences of method calls} to reproduce failures~\cite{Jin:BugRedux:ICSE:2012}, detect malicious behaviors~\cite{Gorji:MalwareDetection:ACMSE:2014}, debug applications~\cite{Murtaza:FaultLocalizationTheory:GTSE:2015}, profile software~\cite{Elbaum:Profiling:TSE:2005}, optimize applications~\cite{Zhao:Optimization:OOPSLA:2014}, and mine models~\cite{Mariani:Revolution:ISSRE:2012,Mariani:DynamicAnalysis:TSE:2011,7927993}. 
Unfortunately, extensively recording sequences of function calls might introduce an annoying overhead and cause unacceptable slowdowns, as for example experienced by Jin et al.~\cite{Jin:BugRedux:ICSE:2012}. Slow software is a major threat to the success of a project, indeed it is reported as one of the main reasons why users stop using applications~\cite{Ceaparu2004,Miller1968,Hoxmeier00systemresponse}.

Since preventing any interference with the user activity is a mandatory requirement, in many cases monitoring techniques have to limit the amount of collected data. This can be achieved in several ways: by limiting the portion of the system that is monitored~\cite{Pavlopoulou:ResidualCoverage:ICSE:1999,apiwattanapong2002selective,orso2005selective}; by distributing the monitoring activity among multiple instances of a same application running on different machines~\cite{Bowring:MonitoringDeployedSoftware:PASTE:2002,Orso:GammaSystem:ISSTA:2002}; by collecting events probabilistically~\cite{Bartocci:Adaptive:ICRV:2012,Jin:Sampling:SIGPLAN:2010,chilimbi2009holmes,Liblit:BugIsolation:SIGPLAN:2003}; and by collecting bursts of events rather than full executions~\cite{Hirzel:BurstyTracing:FDDO:2001,prepost}. Limiting the amount of collected data 
reduces the effectiveness of the approaches that must work with a limited number of samples. 

Since non-negligible overhead levels can be hardly recognized by users as long as the overhead lasts for few interactions~\cite{NIER17,Cornejo:FunctionCallMonitoring:JSS:2020,killeen_optimal_1987}, monitors could be feasibly used to collect fairly complete traces for a limited amount of time. In particular, a monitor might be turned on and off several times during a program execution in order to collect \textit{bursts}, that is, chunks of executions with no internal gaps~\cite{Hirzel:BurstyTracing:FDDO:2001}. Since the monitor is used intermittently, its impact on the user experience is limited. 




Unfortunately, individual bursts capture only part of the history of an execution, providing scattered evidence of the behavior of the monitored software. To \begin{mod}obtain additional information while controlling the overhead, \end{mod}this paper \begin{mod}proposes \Controlled (CBR), a novel monitoring technique for recording bursts whose activation and deactivation is controlled by the operations performed by the users on the target application.\end{mod}
In particular, when a new user \begin{mod}operation is started,\end{mod}
the monitor that records the burst is activated, and once the user request has been fully processed, the burst is finalized and the monitor is turned off. 

Since the recording of the burst is controlled, \CBR can also recombine the recorded bursts a-posteriori to obtain a comprehensive picture of the behavior of the monitored software. In particular, \CBR annotates bursts with state information that captures the state of the monitored application at the beginning and at the end of each burst (i.e., before and after a user 
\begin{mod}operation\end{mod}
 is performed). This information allows \CBR to generate a hierarchical finite state model that represents the behavior of the monitored application when used in the field. In this work we refer to bursts composed of sequences of method calls due to their wide applicability, but the same concepts can be applied to bursts that include other information.



We evaluated \CBR on ArgoUML and found that \CBR provides a valuable tradeoff between overhead and data accuracy compared to regular sampling techniques that probabilistically record bursts without controlling when the execution of the burst begins and ends. 

This paper is organized as follows. 
Section~\ref{sec:CBRFramework} describes \CBR. Section~\ref{sec:empirical_cbr} presents the empirical validation we conducted to assess our technique. Sections~\ref{sec:cbr_results} reports the empirical results. Finally, sections~\ref{sec:related} and~\ref{sec:cbr_discussion} discuss related work and provide concluding remarks, respectively.




\section{CBR Framework}\label{sec:CBRFramework}


\begin{figure}[t]
  \centering
    \includegraphics[width=\columnwidth]{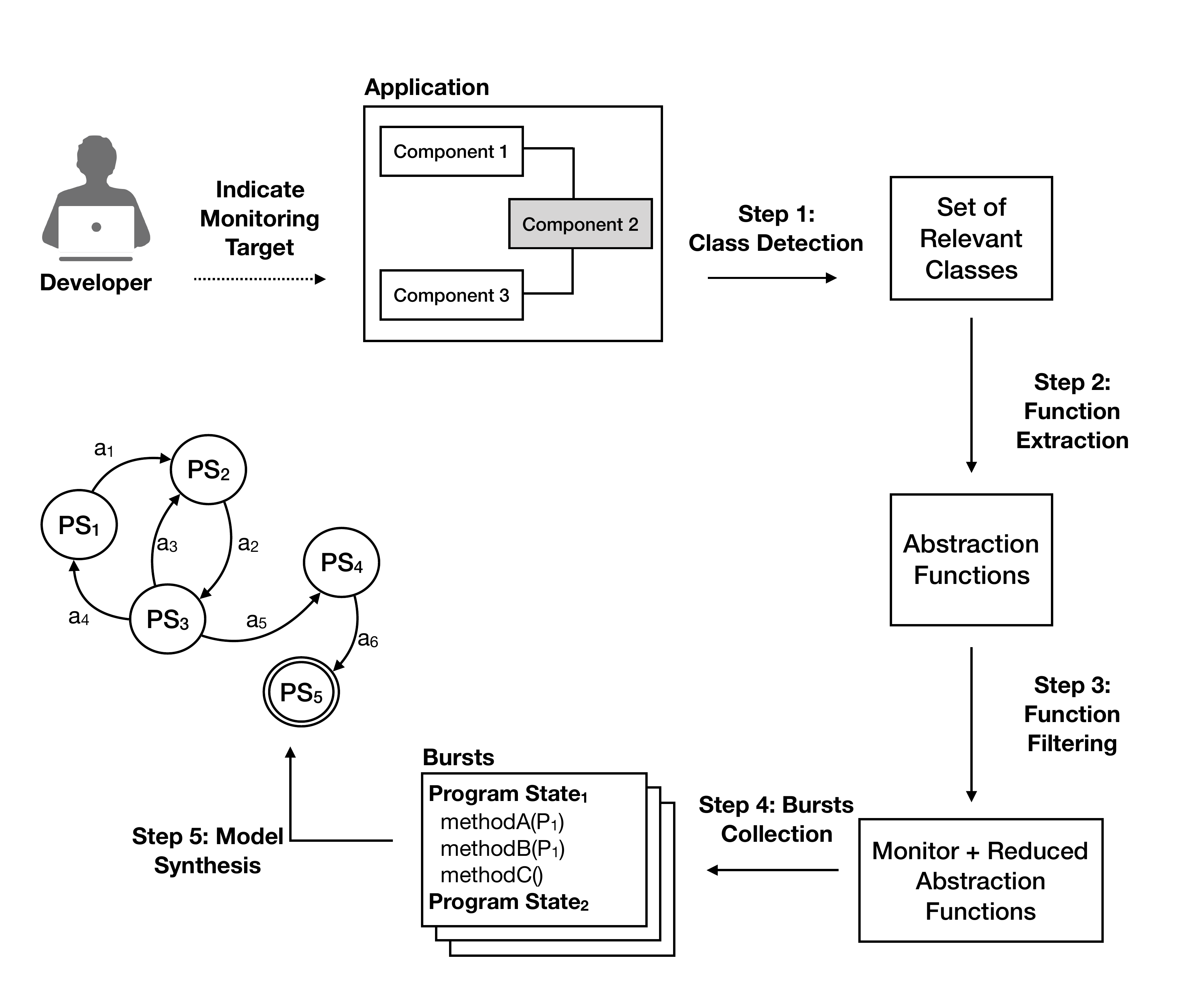}
    \caption{Overview of \Controlled approach.}
    \label{fig:CBR}
\end{figure}

\Controlled (\CBR) is a technique that can reconstruct an approximate representation of the behavior shown in the field by a program and represents it as a hierarchical finite state model. The resulting model is approximated because its states and transitions are obtained by heuristically combining the information in the recorded bursts, especially exploiting the state information that appears at the beginning and end of each burst. 
Figure~\ref{fig:CBR} shows the steps that compose \CBR.

%


The first step of the technique requires that 
the developer specifies the components that must be monitored in the field. Although the full application can be monitored, the developers might be interested in the behavior of some components only. Based on this input, \CBR analyzes the code of the application and identifies the scope of the instrumentation that must be added to the program, which includes all the classes that may directly or indirectly affect the behavior of the monitored components (step 1). We refer to the identified classes as the \emph{relevant classes}.  


To associate contextual information useful to generate the final finite state model with the recorded data, \CBR records bursts that both start and end with state information. The state information should be accurate enough to distinguish the different logical states of the components, and inexpensive enough to be obtained at runtime without introducing a significant overhead. 
To identify a small but relevant amount of state information to be recorded with bursts, \CBR \emph{automatically} derives the \textit{Abstraction Functions} 
that can be used at runtime to produce abstract representations of the concrete program states (step 2). The intuition is that the values of the state variables are relevant only to the extent they can affect the actual execution of the program, and this largely depends on the conditions that are computed on the state variables. For this reason, \CBR analyzes the set of relevant classes looking for Boolean functions that can capture the behavior of the program. For example, if a relevant class implements an operation \texttt{op} whose code checks if the state variable \texttt{x} is positive, \CBR will output the \texttt{x>0} abstraction function, which is assumed to capture a relevant state property. This is intuitively true by construction since it is a condition that is known to influence a computation in one of the relevant classes, that is, the execution takes different paths in \texttt{op} depending on the value of \texttt{x}. Note that only the satisfaction of the condition is relevant, while the specific value of \texttt{x} is not relevant for the computation. This is why \CBR embeds the evaluation of the conditions and not the values of the program variables in the abstract program states.
\CBR obtains the abstract representation (i.e., an abstract state) of a concrete state by applying all the available abstraction functions (i.e., conditions), thus producing a set of function evaluations.




Once \CBR has derived the abstraction functions from all the relevant classes, it analyzes and compares the resulting functions to eliminate the ones that are redundant (e.g., an abstraction function defined as the negation of another abstraction function adds no information about a concrete state of the program). Reducing the set of abstraction functions is useful to improve the performance of the approach, which has to compute fewer functions at runtime and to save smaller abstract states. The set of the remaining functions are embedded into the software monitor that is used at runtime to collect bursts decorated with state information (step 3). 


In the field, \CBR collects bursts coherently with computations (step 4), that is, every time the application starts \begin{mod}a new user operation, a burst is collected with a given probability. A user operation starts with a user input (e.g., a click or a keyboard input) and ends with a feedback to the user (e.g., a result shown on the GUI)\end{mod}. When the burst is recorded, the resulting trace includes a representation of the abstract program state, the sequence of the executed methods, and again a representation of the abstract program state reached at the end of the computation caused by the user \begin{mod}operation\end{mod}. 

The collected bursts are finally analyzed offline. The abstract program states at the beginning and at the end of each burst 
abstractly represent the state of the system at the time the trace 
was recorded. To summarize the observed executions, \CBR creates a Finite State Model (step 5), where each abstract state that occurs in the bursts is a different state of the model and each user \begin{mod}operation\end{mod} that causes a transitions between two abstract states is represented with a transition of the model. The sequences of method calls that can be executed as a consequence of a user \begin{mod}operation\end{mod} are represented as annotations of the transitions. 

The resulting model thus captures the activity observed in the field at multiple abstraction levels \begin{mod}(for this reason it is a hierarchical model)\end{mod}: the states and transitions show how the operations performed by users affect the status of the system, while the annotations show the actual methods involved in the computations.


In the next section, we introduce a running example that we use in the paper to present the steps of the approach.

\subsection{Running example}

Let us consider a Java program that includes the \texttt{Cart} class (see Listing~\ref{code:cart_class}), which represents a simplified shopping cart of an online store. The \texttt{Cart} class implements four different methods and includes the \texttt{Product} class: 
the \texttt{addItem} method adds a \texttt{Product} object to the shopping cart, 
the \texttt{emptyCart} method empties the shopping cart,
the \texttt{calculateTotal} method estimates the total price of all the products in the cart, and
the \texttt{applyDiscount} method applies a discount to the total price.

\JCODESTEP{code:cart_class}{Shopping Cart Class}{Cart.java}{1}

Let us assume that the developer wants to collect data about how a certain program uses the \texttt{Cart} class. 
In the following sections, we discuss how \CBR can address this scenario. Of course, here we consider the monitoring of a single class in a small program for simplicity. In reality,  multiple classes and packages can be selected as targets.

\subsection{Step 1: Class Detection}

The \textit{Class Detection} step identifies the set of classes that may directly or indirectly determine the sequence of execution of the methods in the target components specified by the developer. These classes are automatically identified by first computing the dependency graph of the classes in the program. Then \CBR transitively follows the dependency edges in the graph starting from the target classes. Every class reached during this process is included in the set of the relevant classes analyzed in the second step of the process. 

In the running example the process is quite simple and both the \texttt{Cart} and the \texttt{Product} classes are selected.

%

\subsection{Step 2: Function Extraction}\label{sec:SD}

\CBR analyzes the relevant classes identified with step 1 to extract the abstraction functions, which capture \emph{conditions} that represent how the values of the state variables may influence the execution of the monitored program. The abstraction functions provide a natural and efficient abstraction mechanism that can be applied at runtime to generate an abstract representation of a program state.
%

To generate the abstraction functions, we use symbolic execution~\cite{Baldoni:2018:SSE:3212709.3182657,Braione:Enhancing:FSE:2013}. The main intuition is that each method in each relevant class can be executed symbolically to derive the \emph{path conditions} corresponding to all the paths that can be executed up to a given bound. Each path condition represents a set of conditions over the state variables and the inputs of the program that may drive the execution toward a specific computation (i.e., towards a specific path from the entry of the method to an exit point). For example, the condition \texttt{Cart.products.length==0} identifies a specific path of the program in the \texttt{calculateTotal} method (the path that does not enter in the \texttt{for} loop at line 33) and can be evaluated at runtime to distinguish the state that may lead to one path rather than another when the \texttt{calculateTotal} method is executed. 

When deriving path conditions from methods, the conditions may include clauses containing the input parameters of the processed methods. Since these path conditions are used to derive abstraction functions that might be computed at any time of the execution of a program, and not necessarily when methods with specific parameters are invoked, these clauses are removed from the path conditions when turning them into abstraction functions. For instance, the abstraction functions resulting from the symbolic execution of the methods in the \texttt{Cart} class are shown in Table~\ref{table:abstractfunctions}.






\begin{table*}[ht]
\scriptsize
\centering
\caption{Abstract Functions extracted from \texttt{Cart} class example.}
\label{table:abstractfunctions}
\begin{tabular}{lll}
\toprule
\textbf{Identifier}          & \textbf{Code ref.} & \textbf{Abstract Functions}                                \\
\midrule
addItem - Function 1        & Line 13  & \texttt{Cart.nProducts != 0 \&\& Cart.products.length \textgreater{}= 0}                                                           \\
addItem - Function 2        & Line 14  & \texttt{Cart.nProducts == 0 \&\& Cart.CART\_SIZE \textgreater{}= 0 \&\& Cart.nProducts \textless Cart.CART\_SIZE}                           \\
addItem - Function 3        & Line 14  & \texttt{Cart.nProducts == 0 \&\& Cart.CART\_SIZE \textgreater{}= 0 \&\& Cart.nProducts \textgreater{}= Cart.CART\_SIZE}                     \\
addItem - Function 4        & Line 13  & \texttt{Cart.nProducts == 0 \&\& Cart.products.length == 0}                                                                                 \\
applyDiscount - Function 1  & Line 26  & \texttt{Cart.nProducts \textgreater 0 \&\& Cart.products.length \textgreater 0}                                                             \\
applyDiscount - Function 2  & Line 26  & \texttt{Cart.nProducts \textgreater 0 \&\& Cart.products.length == 0}                                                                       \\
applyDiscount - Function 3  & Line 27  & \texttt{Cart.nProducts \textgreater 0 \&\& Cart.products.length \textgreater 0 \&\& Cart.products.{[}0{]}.value \textgreater{}= Cart.PRICE} \\
applyDiscount - Function 4  & Line 27  & \texttt{Cart.nProducts \textgreater 0 \&\& Cart.products.length \textgreater 0 \&\& Cart.products.{[}0{]}.value \textless Cart.PRICE}       \\
calculateTotal - Function 1 & Line 35  & \texttt{Cart.products.length \textgreater 0 \&\& Cart.products.{[}0{]}.taxFree == true}                                                     \\
calculateTotal - Function 2 & Line 37  & \texttt{Cart.products.length \textgreater 0 \&\& Cart.products.{[}0{]}.taxFree == false}                                                    \\
calculateTotal - Function 3 & Line 33  & \texttt{Cart.products.length \textgreater 0}                                                                                                \\
calculateTotal- Function 4  & Line 33  & \texttt{Cart.products.length == 0}                                                                                                          \\
emptyCart- Function 1       & Line 21  & \texttt{Cart.nProducts \textless{}= 0}                                                                                                      \\
emptyCart- Function 2       & Line 21  & \texttt{Cart.nProducts \textgreater 0 \&\& Cart.CART\_SIZE \textgreater{}= 0} \\
\bottomrule                                                             
\end{tabular}
\end{table*}

When an abstraction function is evaluated at runtime, it is evaluated against the state of the program. \CBR uses a monitor to intercept the creation and the destruction of the monitored objects of the program so that functions can be evaluated efficiently on the existing objects. A function evaluates to true if it exists a set of objects in the program state that can satisfy it. For instance, the first abstraction function in Table~\ref{table:abstractfunctions} evaluates to true if there exists a cart with at least one element.

In addition to true and false values, an abstraction function can also produce the value \emph{unknown}. This happens when some of the elements that appear in the abstraction function cannot be evaluated. For instance, the clause \texttt{Cart.products.[0].taxFree == true} evaluates to \emph{unknown} if there is no element at the position 0 of the array. If a clause evaluates to unknown, the full abstract function returns unknown.

The result of the evaluation of a set of abstraction functions is an array of ternary values, referred as the \emph{Abstract Program State}. More formally, given a program $P$ with a concrete state $S$, its abstract program state $abs(S)= \{v_i | v_i = f_i(S), \forall f_i \in AF\}$, where $AF$ is \begin{mod}an ordered\end{mod} set of the available abstraction functions. \begin{mod}Note that the specific ordering is not important, but it is important to keep it consistent across all the evaluations so that the abstract states are comparable, that is, the value $v_i$ in an abstract state must be produced by the same function $f_i$ every time the abstract state is computed.\end{mod}

\subsection{Step 3: Function Filtering}\label{sec:monitor}

Symbolic execution normally produces a large number of path conditions since it considers every possible execution path for every method under analysis up to a given bound. Turning all the resulting path conditions into abstraction functions to be used in the field would not be practical and would cause unacceptable overhead levels. 
For this reason, \CBR filters out the least useful path conditions returned by symbolic execution, to guarantee a good compromise between the accuracy of the state information that is traced and the cost of producing such a state information.

From the original set of path conditions, \CBR just needs the ones that give unique information about the behavior of the application, that is, information that is not subsumed by the information provided by the evaluation of the other functions. 
To find the optimal set of conditions to be used as abstraction functions, \CBR assesses the available conditions against a set of test executions and discards the ones that do not contribute in distinguishing the program states. 

\begin{figure}[h]
\centering
\includegraphics[width=0.5\columnwidth]{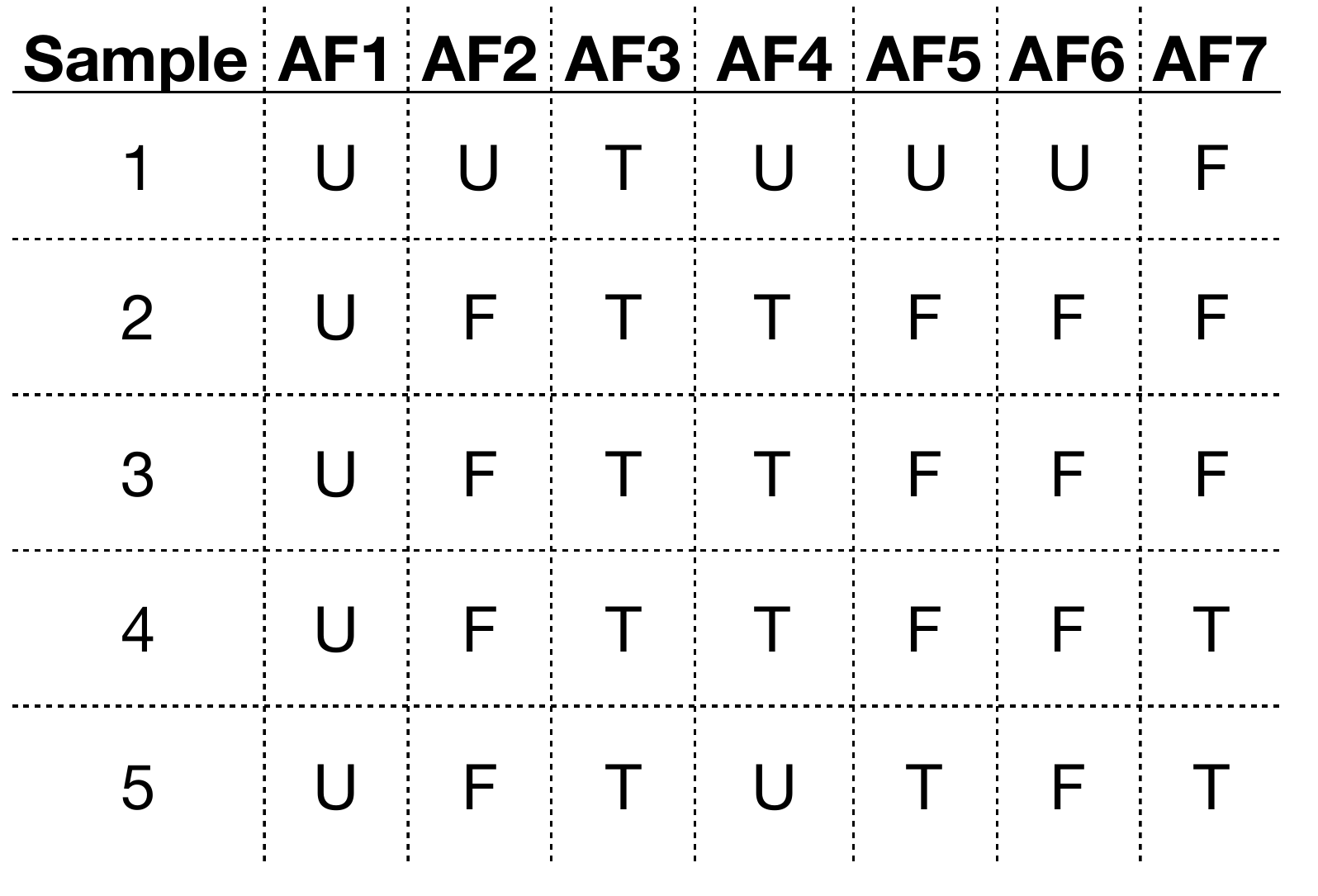}
\caption{Sample matrix with 5 evaluations of 7 abstraction functions.}
\label{fig:matrix}
\end{figure}

In particular, \CBR executes the monitored program covering diverse scenarios (e.g., using a set of system test cases) and evaluates all the available abstraction functions every time a method of the program is invoked. This produces a large number of evaluations for all the available abstraction functions. The collected data can be represented in a matrix where the abstraction functions appear on the columns and the evaluations on the rows. Each row corresponds to an evaluation of all the abstraction functions performed at the time a method was invoked (the specific method is not relevant). On the other hand, each column includes all the values that have been returned by an abstraction function across all the function evaluations that have been performed. Each cell may evaluate to true (T), false (F), or unknown (U). Figure~\ref{fig:matrix} shows a sample matrix with 7 abstraction functions evaluated 5 times.

\begin{figure}[h]
\centering
\includegraphics[width=0.9\columnwidth]{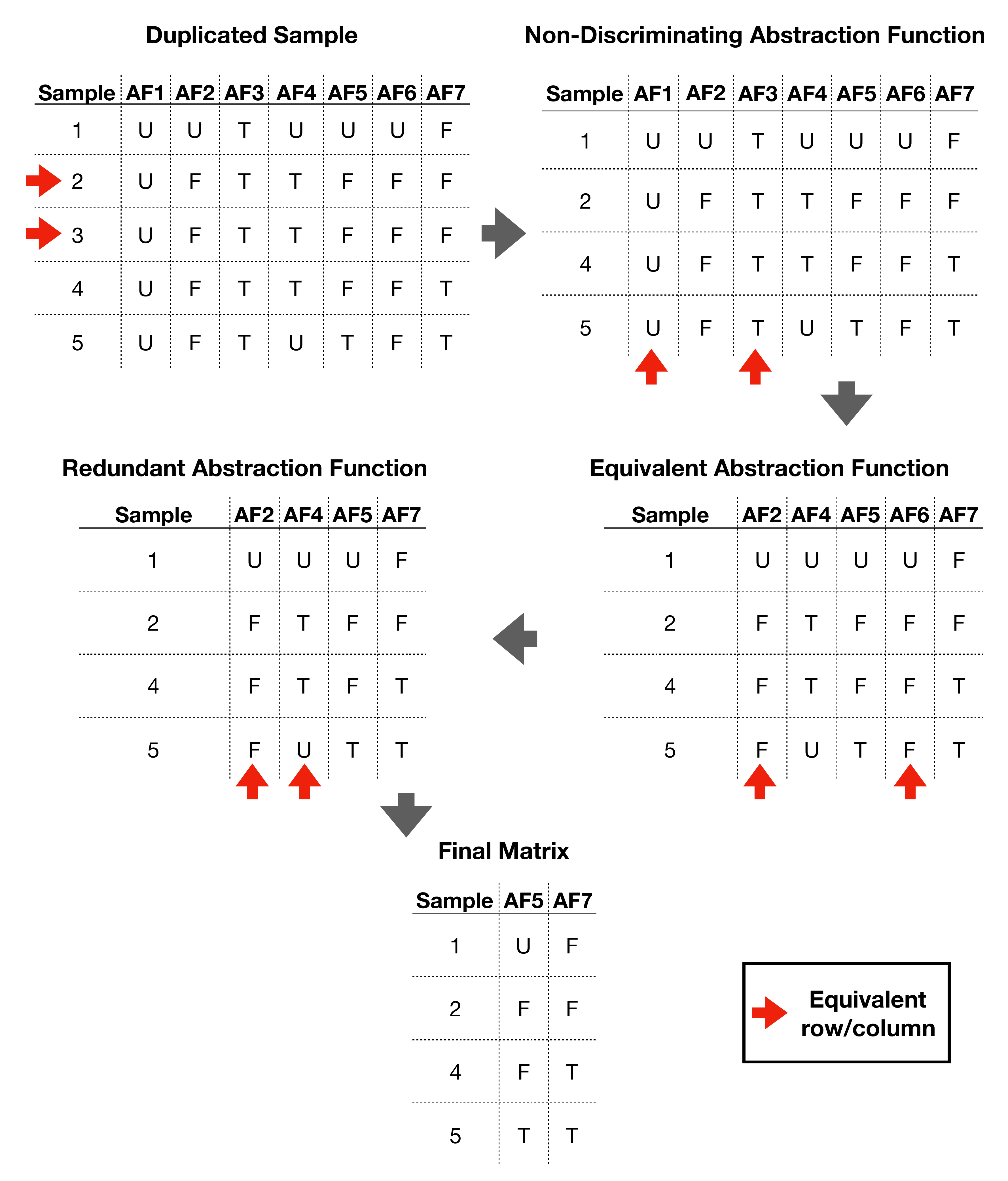}
\caption{Example filtering process.}
\label{fig:pc_filtering}
\end{figure}

The filtering process goes through four steps that reduce the size of the matrix, until reaching a small number of abstraction functions that are really indispensable to generate the abstract program states. The four steps are: \textit{Removal of Duplicated Samples}, \textit{Removal of Non-Discriminating Abstraction Functions}, \textit{Removal of Equivalent Abstraction Functions}, and finally \textit{Removal of Redundant Abstraction Functions}. Figure~\ref{fig:pc_filtering} shows the four steps applied to the matrix in Figure~\ref{fig:matrix}. We describe the four steps below.

\smallskip

\textit{Removal of Duplicated Samples}: this step reduces the size of the matrix by removing duplicated rows, which are useless for the purpose of determining the ability of the functions to distinguish different concrete states. In addition to increase the efficiency of the next steps, since the matrix becomes smaller, this step is necessary to facilitate the identification of the redundant functions (see last rule). Formally, given the matrix $M_{ij}$ with $i=1 \ldots K$, $j=1 \ldots N$, two rows $M_{i_{1}}$ and $M_{i_{2}}$ are duplicated if $M_{i_{1}j} = M_{i_{2}j}$ for all $j = 1 \ldots N$.
In our example, the second and third rows are duplicated and thus only one of them is preserved for the rest of the analysis. 

\smallskip

\textit{Removal of Non-Discriminating Abstraction Functions}: this step eliminates those abstraction functions that never change their values throughout all the evaluations, being de facto constant, and thus not contributing to distinguishing the program states at runtime. These functions can be for instance the result of the analysis of infeasible program paths. More formally, an abstraction function in column $J$ is non-discriminating if $M_{1J} = M_{2J} = \ldots = M_{kJ}$.
In our example, the abstraction functions AF1 and AF3 are non-discriminating and are thus removed from the matrix.

\smallskip

\textit{Removal of Equivalent Abstraction Functions}: this step eliminates abstraction functions that consistently return the same abstract values. Indeed, keeping only one of this function is enough. More formally, two functions in columns $j_1$ and $j_2$ are equivalent if their columns are the same, that is, $M_{ij_1} = M_{ij_2}$ for all $i=1...k$. In the example, AF2 and AF6 are equivalent, thus AF6 is dropped from the matrix.

\smallskip

\textit{Removal of Redundant Abstraction Functions}: this step 
removes the abstraction functions that are not needed to actually distinguish the possible program states. The process works by removing one abstraction function at time (i.e., one column at time) and checking if the remaining functions are still sufficient to distinguish all the abstract program states collected so far. If a column is not needed, no rows become equal due to a column dropped from the matrix, that is, the available states can still be distinguished with the remaining functions. Note that after the application of the first rule, only distinct rows remained in the matrix: each row is thus a distinct abstract program state discovered by the tests. If the removal of a column generates two equal rows, it implies that the remaining functions are not sufficient to distinguish the actual states of the program. 
The application of this process to our example causes the removal of functions AF2 and AF4.

\smallskip

The functions that remain at the end of this process (i.e., the remaining columns) are the ones used to produce the abstract states at runtime. In our example, the two remaining functions are AF5 and AF7.
\subsection{Step 4: Bursts Collection}\label{sec:collect_bursts}

\CBR records bursts synchronously with user \begin{mod}operations\end{mod}, to make sure to collect state information when the monitored application is in a sound and steady state. In particular, when the user interacts with the application \begin{mod}asking for an operation, \end{mod}
\CBR decides with a given probability if a burst must be collected. If the burst is collected, the abstract state corresponding to the current concrete state is collected, the actual burst is recorded (e.g., the sequences of methods executed as a consequence of the \begin{mod}request\end{mod}), and when the monitored application has completed the computation and has produced a result, the abstract state is collected again. More formally, a burst $B$ is a tuple $B = \langle label, abs(S_a), trace, abs(S_b) \rangle$, where $label$ is the user \begin{mod}operation\end{mod} that originated the burst, $abs(S_a)$ and $abs(S_b)$ are two abstract states collected before and after the execution of the burst, and the trace $trace$ is the sequence of events (method calls in our case) collected in the field.

While a program is used, a number of bursts $B_1 \ldots B_n$ are collected and used in the final step to reconstruct a representation of the behavior of the program.

%
%
%
%
%
%
%


\subsection{Step 5: Model Synthesis}\label{sec:simulating}

The set of bursts represent chunks of executions collected in different instants. Each burst carries some knowledge about the behavior of the system, that is, how a user 
\begin{mod}operation\end{mod}
made the monitored system to change from a given (abstract) state to another (abstract) state, and the relative internal computation that has been observed (e.g., the sequence of method calls). It is however important to gain a comprehensive picture of how an application or a set of components behave in the field, putting the collected bursts in context. This is something sampling techniques that collect bursts without context cannot do~\cite{Hirzel:BurstyTracing:FDDO:2001}, while controlled burst recording can do thanks to the presence of the state annotations.

\CBR generates a finite state model to capture the knowledge extracted in the field. In particular, it produces a finite state model (FSM) where state transitions are annotated with information about the computations that may happen in the target components when that specific transition is executed.

The way the FSM is reconstructed is driven by the abstract states, that is, each distinct abstract state in the recorded bursts is a different state of the FSM, and the user 
\begin{mod}operations\end{mod}
are transitions in the FSM. The content of the burst produces the annotations.  More formally, an annotated FSM is a tuple $G = (S,T, a)$, where $S$ is a finite non-empty set of states,  $T \subseteq Label \times S \times S $ is a finite set of transitions between states in $S$ with a label in $Label$, and $a: T \rightarrow \mathbb{TRACE}$ is a function that associates each transition in the model with a set of traces ($\mathbb{TRACE}$ denotes the powerset of all the possible traces). Note that the FSM captures various state transitions that have been observed in the field without encoding a notion of initial and final states.

Given a set of bursts $B=\{B_1, \ldots B_n\}$, with $B_i = \langle label_i, abs(S_a)_i, trace_i, abs(S_b)_i \rangle$, the corresponding FSM $(S,T, a)$ is defined as follows.
\begin{itemize}
\item $S=\bigcup_i (abs(S_a)_i \cup abs(S_b)_i)$ is the union of all the states in the bursts,
\item $T=\{(l, s_a,s_b) | \exists b \in B \wedge b= \langle l, s_a, trace, s_b \rangle \}$, is a representation of all the state transitions caused by the user \begin{mod}operations\end{mod} present in the bursts,
\item $a(t)=A$ with $t = (l, s_a,s_b) \in T$ implies $A= \bigcup_{\langle l, s_a, trace_j, s_b \rangle} trace_j$, is a function that annotates each transition with the corresponding set of traces. 
\end{itemize}
 
The model is simply created by sequentially mapping each burst into the corresponding states, transitions, and annotations.

\begin{figure}[h]
\centering
\includegraphics[width=0.70\columnwidth]{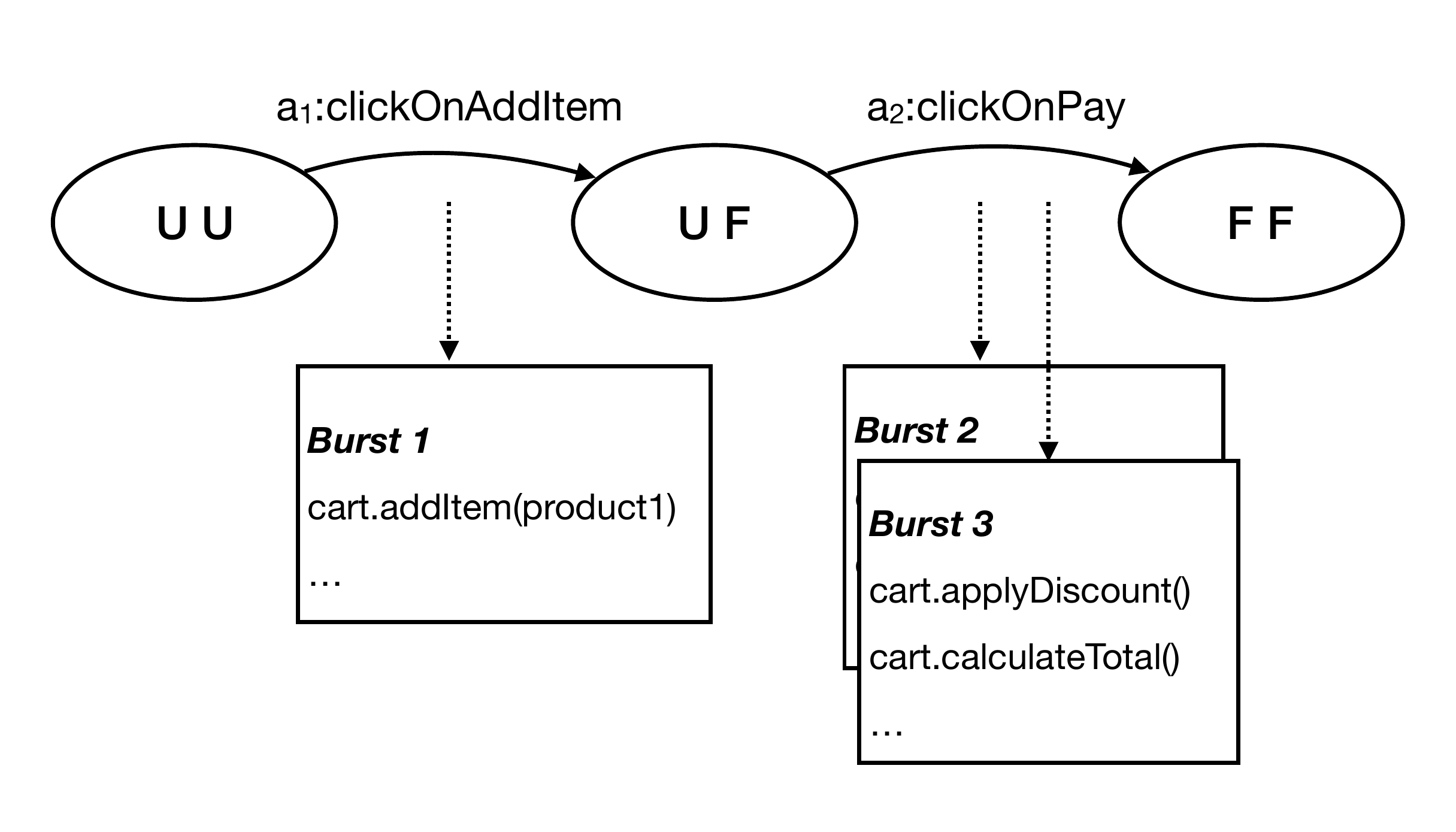}
\caption{Excerpt of FSM derived from a set of bursts.}
\label{fig:burst_plus}
\end{figure}

Figure~\ref{fig:burst_plus} shows an excerpt of the resulting model. Each state of the model is a different abstract state of the program (for simplicity we only report two values in the represented states). The transitions show changes in the current state of the monitored components. Each transition is annotated with traces (as represented with dotted arrows in Figure~\ref{fig:burst_plus}). 

Note that all the steps of the approach are driven by the initial selection made by the user. Thus, both the derived abstraction functions and the resulting model specifically capture the behavior of the monitored components.  

In a nutshell, \CBR provides useful information at two different levels:
(1) it shows how the usage of the application affects the state of the monitored components, and 
(2) it shows the computations that might be produced in the monitored components every time a transition is traversed.

In the next section, we evaluate \CBR in comparison to other approaches for sampling executions from the field.

%
%
%
%
%


\section{Empirical Evaluation}\label{sec:empirical_cbr}

This section presents the empirical evaluation that we conducted for \Controlled, to assess the approach in comparison to  sampling techniques. Note that here, even if \CBR itself record bursts with a given probability, we use the term sampling techniques to refer to the techniques that collect bursts in an uncontrolled way (i.e., not synchronously with the processing of user \begin{mod}operations\end{mod}).
In particular, we aim to answer the following research questions.

\begin{itemize}
	\item \textbf{RQ1: What is the overhead introduced by \Controlled?} This research question studies the overhead introduced by \CBR in comparison to other sampling techniques.

	\item \textbf{RQ2: What is the precision of the model generated by \Controlled?} This research question investigates the precision of the information captured in the model produced by \CBR with respect to the behaviors shown by the system. Since sampling techniques do not recombine the recorded traces, their precision is always 1 and do not need to be studied empirically.

	\item \textbf{RQ3: What is the completeness of the model generated by \Controlled?} This research question studies the capability of \CBR to capture the actual behaviors shown by the system, in comparison to sampling techniques.
\end{itemize}

\subsection{Prototype}
Our tool is implemented in Java and targets Java programs. It integrates third-party tools for static analysis, symbolic execution, and monitoring. In particular, \CBR uses WALA~\cite{WALA} to statically analyze the code of the monitored program and identify the relevant classes. In addition, it uses JBSE~\cite{Braione:Enhancing:FSE:2013} to symbolically analyze the relevant classes and produce the path conditions that are transformed into abstraction functions. 
The generation of the conditions from each analyzed method is bounded by limiting the number of branches and the number of states that can be traversed sequentially to 10 and 1000, respectively. The analysis time of each method is also limited to 60 seconds.
Finally, \CBR uses AspectJ~\cite{AspectJ:website:2018} to collect data about the executed methods. 

\subsection{Experimental Subject}

To empirically answer the three research questions, we selected ArgoUML~\cite{ArgoUML:website:2018}, which is a non-trivial (389,952 locs) open source Java application for editing UML diagrams that can be used in a variety of ways (e.g., to produce largely different diagrams) and whose behavior can be studied in terms of the executions produced in the field.  

To consider the situation in which 
\begin{mod}we are\end{mod} interested in a specific part of the application, we selected the entire  \begin{small}\texttt{activity}\end{small} package as the target of the monitoring activity. This package manages all the functionalities related to the design and management of activity diagrams in ArgoUML. 
As monitoring objective, we consider collecting calls to methods implemented in the classes of the target package, including the values of the parameters. In order to recreate proper executions of ArgoUML, we have \begin{mod}implemented 25 test cases reproducing typical usage scenarios for the \begin{small}\texttt{activity}\end{small} package. Each test case consists of drawing a different activity diagram using different elements and features. \end{mod}
To avoid non-determinism and for the reproducibility of the test cases, we 
\begin{mod}recorded and executed\end{mod} them with the Sikulix testing tool~\cite{sikulix}. All the measurements have been obtained by repeating the execution of the tests three times. 

The test cases, our tool, and the experimental material are available in the following repository \url{https://github.com/cbr-paper/CBR_Experimental}.

\subsection{Function Filtering}

In this section, we discuss the result of the filtering process applied to our experimental subject (step 3). When running \CBR in the considered setting, the symbolic executor produced 5,732 abstraction functions. \CBR then identified and filtered out all the \textit{Non-Discriminating}, \textit{Equivalent}, and \textit{Redundant} Abstraction Functions. Overall, \CBR reduced the number of abstraction functions to be used to generate the abstract program states from 5,575 to 157 abstraction functions, filtering out 97.2\% of the functions. 
Table~\ref{table:filtering} shows the number of functions that have been filtered out in each step by \CBR, confirming the usefulness of all the steps.

\begin{table}[h]
\centering
\caption{Function Filtering}
\label{table:filtering}
\scriptsize
\begin{tabular}{lc}
 & \# Abstraction Functions \\ 
\toprule
Initial number of abstraction functions & 5,732 \\ 
\midrule
Non-Discriminating Abstraction Functions & 1,631\\
Equivalent Abstraction Functions & 2,646\\
Redundant Abstraction Functions & 1,298\\
\midrule
Final set of Abstraction Functions & 157\\
\bottomrule
\end{tabular}
\end{table}

\subsection{Experiment Design}\label{sec:cbr_exp_design}

In our assessment, we compare \CBR to sampling techniques, which have been used widely for monitoring applications in the field~\cite{Jin:Sampling:SIGPLAN:2010,liblit2005scalable,joshi2017runtime,BabaeeGF18}. These solutions limit the overhead by collecting data  with a given probability. 
Note that differently from \CBR, these bursts have neither a semantics associated with the processing of the user requests nor the state annotations, but consist of traces containing a fixed number of method calls. 
We considered sequences of length $30$, similarly to the configuration setup \begin{mod}used for \textit{Bursty Tracing}~\cite{Hirzel:BurstyTracing:FDDO:2001}, which is the closest technique to \CBR.\end{mod}

In our evaluation, we compare \CBR to Sampling monitoring configured with two sampling probabilities: $5\%$ and $10\%$ (higher sampling frequencies might be hardly considered since they may interfere with the user activity). 
We thus compared three solutions: \CBR, Sampling monitoring with 5\% sampling probability, and Sampling monitoring with 10\% sampling probability.

\smallskip

To answer \emph{RQ1}, we measure the overhead introduced in the subject program by the three monitoring techniques when running the available test cases. To measure overhead, we collect the system response time of each user operation that is executed. 


To capture how monitoring may impact the execution of operations of different nature, we organize the data based on the well-known and widely accepted classification from the human computer interaction area proposed by Seow~\cite{seow}. In this classification, operations are organized according to four categories, which have been derived from direct user engagement. Each category represents a different type of operation and has a foreseen maximum system response time. The four categories are: 

\begin{itemize}
\item \emph{Instantaneous}: these are the simplest operations that can be performed on an application, such as entering inputs
or navigating throughout menus. Their system response time is expected to be 200ms at most.
\item \emph{Immediate}: these are the operations that generate acknowledgments or very simple outputs. Their system response time is expected to be 1s at most.
\item \emph{Continuous}: these are operations that produce a result within a short period of time to not interrupt the
dialog with the user. Their system response time is expected to be 5s at most.
\item \emph{Captive}: these are operations requiring some relevant processing for which users will wait for results. Their system response time is expected to be 10s at most. These operations are not present in ArgoUML.
\end{itemize}

We study the impact of the monitoring activity per category because the relative overhead can be quite different for each type of operation.

\smallskip

%

To answer \emph{RQ2}, we measure to what extent the behaviors represented in the model generated by \CBR correspond to actual (original) traces of the monitored program (we collected the complete traces of execution, with all the generated function calls, once at the beginning of the experiment). We study this aspect locally to each node in comparison to the actual set of observed behaviors because nodes are the joint points between bursts, which may introduce imprecision. 
That is, we evaluate precision locally (i.e., \textit{node precision}) and then globally (i.e., \textit{overall nodes precision}). In particular, we first assess if the local decisions taken in each node have a corresponding evidence in the original traces, and then we compute a global metric as mean of the local precision of each node.  


\begin{figure}[h]
\centering
\includegraphics[width=0.35\columnwidth]{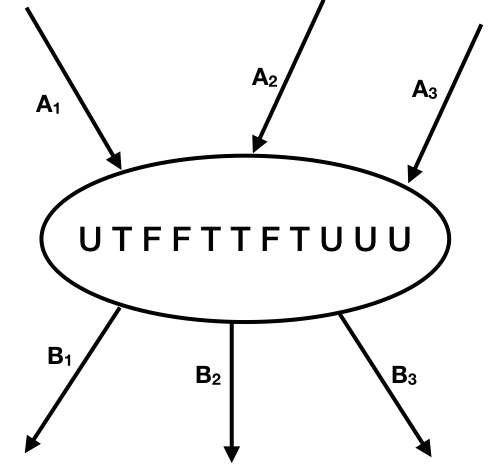}
\caption{\FSA node with state information and incoming and outgoing transitions.}
\label{fig:node}
\end{figure}

To assess \textit{node precision}, for each node in the model we check if every possible sequence of events of length 2 which involves that node is an actual program behavior. 
For example, for the node in Figure~\ref{fig:node} we first generate all the possible combinations of sequence of actions, that is, $A_1 \rightarrow B_1$, $A_1 \rightarrow B_2$, \ldots, $A_3 \rightarrow B_3$, then we verify if the traces produced by these sequences are present in the set of original recorded traces. 

In particular, the node precision for a particular node $node_i$ is computed with the formula $precision(node_i) = CS_i\mathbin{/}TS_i$ where $CS_i$ is the number of sequences of length 2 that traverse $node_i$ and that have a  counterpart in the original traces, and $TS_i$ is the total number of sequences of length 2 that traverse node $i$.
The \textit{overall nodes precision} is obtained by computing the mean \textit{node precision} of all the nodes in the model. For simplicity in the rest of the document we refer to \textit{overall nodes precision} simply as \textit{precision}.

Note that we do not compute precision for sampling techniques because they do not provide any form of generalization of the collected traces, and thus the precision of the extracted information is always 1.

\smallskip

To answer \emph{RQ3}, we measure \textit{trace-level recall} that is, we measure how complete the extracted information is with respect to the complete traces of executions produced by the monitored application. Note that all the monitoring techniques collect only a subset of these traces when the sampling is activated: both \CBR and sampling techniques collect bursts while missing the rest of the computation.  
However, the information that is not recorded from an execution can still be recorded in future executions, thus finally obtaining a more complete picture of the behavior observed in the field.

To measure the ability of the monitoring technique to extract complete traces from the bursts collected from the field, we measure the trace recall for a given trace as  $recall(trace_j) = E_j\mathbin{/}EO_j$ where $E_j$ is the number of calls captured by the monitoring technique, and $EO_j$ is the overall number of method calls in the original trace. We use the available test cases to obtain the program traces. Note that while sampling techniques extract sequences of fixed length, \CBR can reconstruct longer executions combining multiple samples thanks to the state information attached to bursts. In the rest of the document we refer to \textit{trace-level recall} simply as \textit{recall}.

\smallskip

Let us remark that both \CBR and sampling techniques are used to reconstruct information about behaviors observed in the field, and not the general behavior of an application. For this reason, both the precision and the recall metrics are defined with respect to the full traces produced by running the test cases of the monitored application, and not about all the feasible behaviors of the monitored program.


\section{Results}\label{sec:cbr_results}

In this section we present the results of the experiment we conducted to validate \CBR. All the experiments were executed on a computer running macOS version 10.13.6 with a 3.1 GHz Intel Core i7 processor and 16 GB of RAM.


\subsection{RQ1: Performance}

To answer the research question \textbf{RQ1: What is the overhead introduced by \Controlled?} we assessed the impact of the monitors, collecting data about their overhead. To this end, we repeated the execution of our 25 test cases 3 times obtaining 3,459 measurements of the response time of each operation when the various types of monitors are active. Based on the categorization of the operations presented in the previous sections (we assigned the category to each operation considering the system response time of the application without any monitoring), we collected data from 915 operations in the Instantaneous category, 25 operations in the Immediate category, and 213 operations in the Continuous category. Given the nature of the functionalities tested, we did not identify operations in the Captive category.


\begin{table}[h]
\centering
\caption{Performance results with respect to SRT categories.}
\label{table:performance_results}
\scriptsize
\begin{tabular}{lcccc}
\toprule
  \multirow{ 2}{*} {\textbf{Monitoring Technique}} & \multicolumn{3}{c}{\textbf{Monitoring Overhead {[}\%{]}}} \\ 
\cmidrule(l){2-4}
& \textbf{Instantaneous} & \textbf{Immediate} & \textbf{Continuous} \\ 
\midrule
\CBR & 123.78 & 40.70 & 0.71 \\ 
Sampling (P: 10\%) & 17.03 & 4.65 & 1.37 \\ 
Sampling (P: 5\%) & 7.12 & 2.88 & 0.28 \\ 
\bottomrule
\end{tabular}
\end{table}

The mean overhead introduced by each approach can be found in Table~\ref{table:performance_results}. Note that the overhead produced by \CBR corresponds to the overhead experienced while recording every call executed while processing an operation. On the contrary, sampling techniques record only a subset of the methods executed while processing operations. 


For the operations that require less than a second to execute (i.e., Instantaneous and Immediate), \CBR introduces considerable more overhead than the other sampling techniques. 

The higher overhead introduced for operations that can be completed in less than a second is explained by the need of recording the information about the abstract states before and after each burst. Indeed the absolute overhead is still small: for instance, the slowest Instantaneous operations may go from 200ms (its maximum time from the category) to 450ms with the overhead, which is still a difference that can be hardly recognized by users, as also confirmed in other studies where overhead levels up to 180\% are reported as hard to detect for Instantaneous operations, as long as introduced for a few operations in a row~\cite{NIER17, Cornejo:FunctionCallMonitoring:JSS:2020}. 

On the contrary, \CBR performs well with Continuous operations, where high overhead values may result in noticeable slowdowns for the users. Indeed, the cost of collecting state data is well compensated by the duration of the operations.

In summary, \CBR is more expensive than sampling techniques due to the state abstraction and recording activities performed when a burst is recorded. The relative extra overhead is significant for operations that complete quickly, but still hardly recognizable in terms of the absolute slowdown, while the overhead introduced in longer operations to record state information is compensated by the length of the operation.

\subsection{RQ2: Precision}\label{sec:acc_results}

\begin{figure}[h]
\centering
\includegraphics[width=\columnwidth]{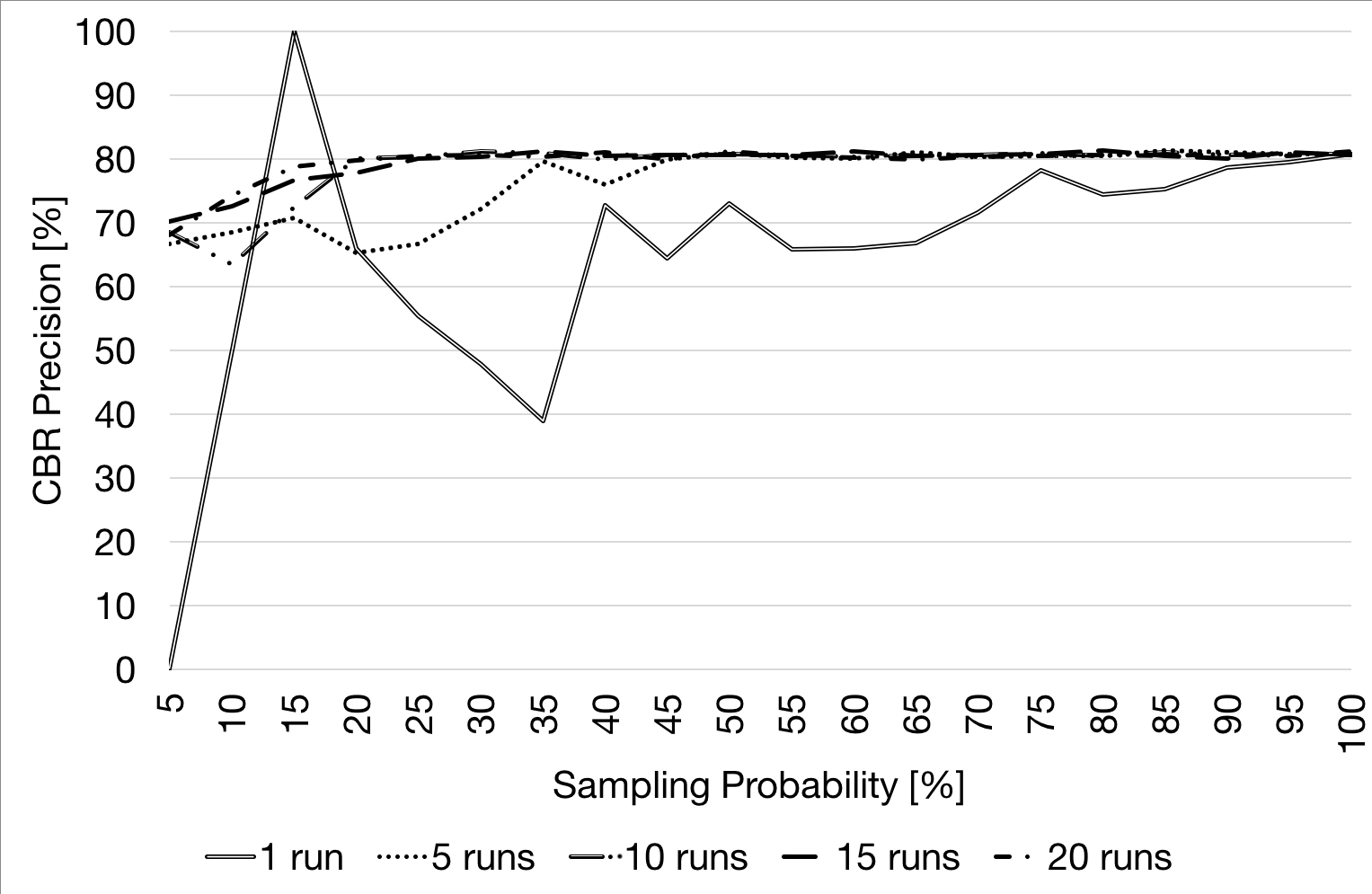}
\caption{Precision results with respect to different runs of the application.}
\label{fig:precision_results}
\end{figure}

For answering the research question \textbf{RQ2: What is the precision of the model generated by \Controlled?} we measured the precision of the information reported in the model generated by \CBR with respect to the set of original traces. 
We only show results for \CBR, since it is the only technique that recombines bursts to obtain a more comprehensive model, possibly introducing imprecision. On the contrary, sampling techniques do not recombine the recorded information, thus the portion of traces collected are always precise, that is, every sequence of method calls recorded corresponds to an actual sequence produced while running the software.

Since \CBR collects a burst with a given probability, we studied how precision changes with respect to the probability to record the bursts and the number of executions of the monitored software. The results are reported in Figure~\ref{fig:precision_results}. 



The ability to recombine bursts produces a loss of precision of about 20\%. Interestingly, \CBR reaches its maximum precision level quite quickly. With the exception of data collected from a single execution, which clearly provides imprecise and unstable information, 15-20 runs with a sampling rate of 20\% are already sufficient to reach the maximum precision. If we reduce the sampling rate, we proportionally need more executions to  obtain the same amount of information. For instance, the same results obtained for 15 runs observed with a probability to collected a burst equals to  20\% could be approximatively obtained with a probability equals to 2\% after having observed 150 runs. Assuming to keep \CBR active in the field for long time, bursts could be feasibly collected with a probability lower than 1\%.

In summary, recombining bursts in a model as \CBR does may introduce imprecision. Based on our preliminary results, \CBR manages to represent combinations of bursts that correspond to actually observed behaviors in 80\% of the cases, while 20\% of these combinations represent behavior not observed in the traces.

\subsection{RQ3: Recall}

\begin{figure}[h]
\centering
\includegraphics[width=\columnwidth]{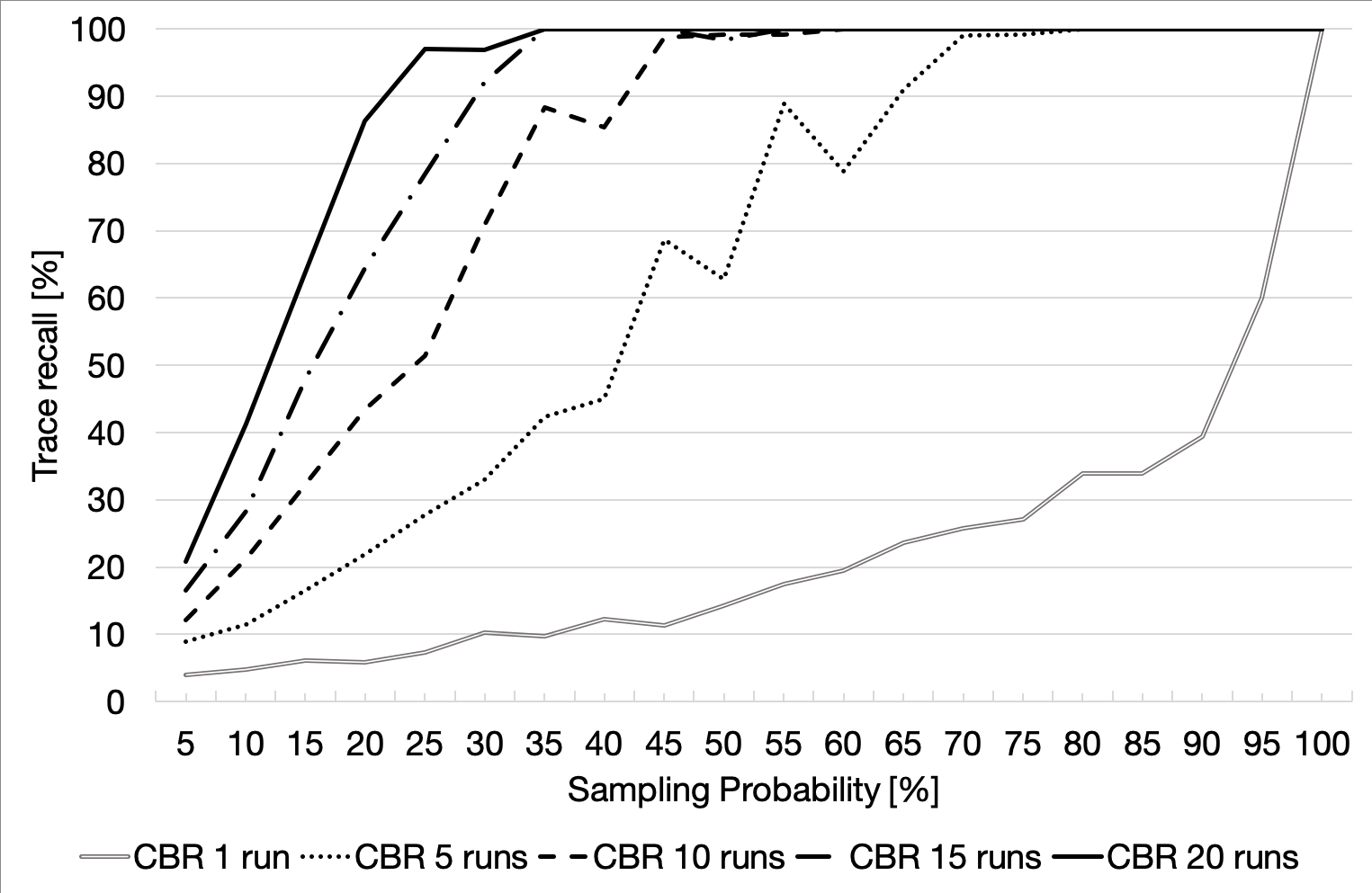}
\caption{Recall results with respect to different runs of the application.}
\label{fig:recall_results}
\end{figure}

In this section, we present the results for the research question \textbf{RQ3: What is the recall of the traces produced by \Controlled?} 

We consider the ability of the various monitoring techniques to capture the behavior observed in the field. Sampling techniques do not have the ability to recombine observations, thus they can capture only short prefixes of the executions that happen in the field. Thus, if sampling has a precision equals to 1 by construction, it has also a nearly 0 recall by construction. In our evaluation, the recall of the sampling techniques ranged from 4.15\% to 7.77\%.

The main objective of \CBR is to obtain a better recall recombining the collected bursts \begin{mod}without annoying the end user\end{mod}. In Figure~\ref{fig:recall_results} we plot how the recall of the model produced by \CBR changes for different sampling probabilities (i.e., different probabilities of recording a burst when a user 
\begin{mod}operation\end{mod}
is performed) and different number of collected executions.



Interestingly, 10 runs of the applications are already sufficient to obtain 90\% recall with a 30\% sampling probability. Again, collecting additional executions allows to reduce the sampling probability still retaining a similar recall, that is, a 3\% sampling probability may approximatively lead to similar recall after the observation of 100 runs. 


%
%


In summary, sampling strategies largely miss the ability to reconstruct the observed executions. On the contrary, \CBR can obtain high recall values by recombining the collected bursts.


\subsection{Threats to Validity and Discussion}
We studied the effectiveness of \CBR in comparison to sampling strategies with a case study based on ArgoUML. We cannot make claims about the generalizability of the results, however, the presented study provides an initial evidence of the complementarity between \CBR and sampling. 
Sampling techniques are useful when collecting extensive evidence of the field behavior is not important, but collecting small but precise evidence is the priority. On the contrary, \CBR can be used when full precision of the extracted information is not a mandatory requirement, and obtaining a comprehensive representation of the observed behavior is more important, for instance to support heuristic program analysis and profiling techniques: \begin{mod}in these cases, \CBR represents a good tradeoff between the introduced overhead, almost not recognizable when using a low sampling rate, the high recall of the collected traces, and the good precision of the reconstructed traces.\end{mod}


\section{Related Work}\label{sec:related}

Our work focuses on the cost-effectiveness of monitoring, aimed at collecting field data in a non-intrusive way, possibly missing a limited amount of information. To this end, we relate \CBR to probabilistic and state-based monitoring approaches.

\textit{Probabilistic monitoring} accounts for lowering the impact of monitoring by collecting runtime information within a certain probability.
Liblit et al.~\cite{Liblit:BugIsolation:SIGPLAN:2003} exploited this strategy to isolate bugs by profiling a large, distributed user community and using logistic regression to find the program predicates that could be faulty. 
Similarly, Jin et al.~\cite{Jin:Sampling:SIGPLAN:2010} presented a monitoring framework called Cooperative Crug (Concurrency Bug) Isolation to diagnose production run failures caused by concurrency bugs. This technique uses sampling to monitor different types of predicates while keeping the monitoring overhead low.
In the same way, Hirzel et al.~\cite{Hirzel:BurstyTracing:FDDO:2001} developed Bursty Monitors, which collect subsequences of events with ad-hoc strategies to construct a temporal program profile.

In general, Probabilistic Monitoring is designed to collect partial information about executions with little ambition to reconstruct a wider picture of the behavior that can be observed in the field, while \CBR focuses on the construction of extended evidence by combining multiple bursts.


\emph{State-based} monitoring approaches focus on using a small subset of variables to represent a significant program state. These techniques are often used either for replaying field executions or for trace analysis (e.g., debugging).

For instance, Orso et al. \cite{orso2005selective} proposed a technique for selectively capturing and replaying program executions. This technique allows to select a subsystem of interest, capture at runtime all the interactions of the applications with its subsystem and then replay the recorded interactions in a controlled environment. For each interaction of the application, the technique captures a minimal subset of the application's state and environment required to replay the execution.
Similar to \CBR, this framework exploits the idea of tracing just the entities defined inside the subsystem of interest. However, this framework is not designed to derive extensive knowledge about the behavior of the system, but focuses on reproduction. 

Diep et al. \cite{Diep:Trace:ISSRE:2008} presented a technique for analyzing traces produced by field applications, in particular to identify and delete irrelevant events from traces that do not offer interesting information for offline analyses.
Before deploying the application, practitioners select a subset of program variables to be used to represent the program state, then while the application is running in the field, the state of these variables is regularly saved before and after each monitored event.
In a second step (i.e., offline time), the technique divides the full trace in several pieces using the variables state as splitting points. After this operation, the technique deletes all the events that do not change the program state and those events that whose occurrence can be re-ordered without affecting the program state, leaving in the trace only the most relevant information for understanding the program behavior.

Contrary to \CBR, this technique does not take into account the monitoring overhead introduced by the action of regularly saving state information, besides the fact that the way the state representation is obtained is specified manually, while  \CBR performs this operation in an automated way.

The heuristic used in \CBR to merge abstract states and produce a finite-state representation of the observed behavior is similar to the one used in some automata inference techniques~\cite{Dallmeier:Adabu:WODA:2006,Marchetto:Ajax:ICST:2008,Mariani:Revolution:ISSRE:2012}. However, none of these approaches are designed to work with bursts as defined in this paper.

Finally, some techniques can be generally useful to optimize resource consumptions and further reduce overhead, such as Delayed Saving that can reduce the cost of persisting the information about the collected events~\cite{Cornejo:DelayedSaving:IEEEAccess:2019}. These kinds of approaches can be integrated in \CBR to further improve its performance. 




\section{Conclusions}\label{sec:cbr_discussion}

\CBR extracts relevant information about the behavior of a software running in the field by collecting bursts of executions synchronously with the computations. That is, each burst corresponds to the processing of a user \begin{mod}operation\end{mod}. Moreover, each burst is annotated with state information collected at the beginning and the end of the burst: this allows \CBR to recombine the information present in the bursts and build a hierarchical model that well captures the behavior observed in the field for a target application. 

We conducted a case study based on Argo UML as preliminary evaluation.
%
%
When assessing performance, we discovered that the overhead introduced by \CBR (in the worst case) with respect to other sampling approaches can be higher for short operations, but similar for the other operations. The higher overhead for short actions is relatively harmless, since it can be hardly recognized by users~\cite{NIER17, Cornejo:FunctionCallMonitoring:JSS:2020}.

When assessing the quality of the extracted information, we discovered that \CBR can reconstruct behavioral information more effectively than sampling techniques, which are highly precise but can hardly produce a fairly comprehensive picture of the observed behaviors. 
Thus, \CBR succeeds in recording useful field data with hardly recognizable impact on the system response time of the monitored application.

As part of future work, we would like to extend the empirical evaluation with other applications and case studies, as well as exploiting the information that can be effectively collected in the field to support program analysis tasks.

\smallskip
\begin{small} \emph{Acknowledgment}.
This work has been partially supported by the EU H2020 ``Learn'' project, which has been funded under the ERC Consolidator Grant 2014 program (Grant Agreement n. 646867) and the ``GAUSS'' national research project, which has been funded by the MIUR under the PRIN 2015 program (Contract 2015KWREMX).\end{small}

\newpage

\bibliographystyle{IEEEtran}
\bibliography{bibliography}

\end{document}